\documentclass{aa}
\usepackage{graphicx}
\usepackage{txfonts}
\usepackage{subcaption}
\usepackage{natbib,twoopt}
\usepackage[breaklinks=true]{hyperref} 
\bibpunct{(}{)}{;}{a}{}{,} 
\makeatletter
\newcommandtwoopt{\citeads}[3][][]{\href{http://adsabs.harvard.edu/abs/#3}%
{\def\hyper@linkstart##1##2{}%
\let\hyper@linkend\@empty\citealp[#1][#2]{#3}}}
\newcommandtwoopt{\citepads}[3][][]{\href{http://adsabs.harvard.edu/abs/#3}%
{\def\hyper@linkstart##1##2{}%
\let\hyper@linkend\@empty\citep[#1][#2]{#3}}}
\newcommandtwoopt{\citetads}[3][][]{\href{http://adsabs.harvard.edu/abs/#3}%
{\def\hyper@linkstart##1##2{}%
\let\hyper@linkend\@empty\citet[#1][#2]{#3}}}
\newcommandtwoopt{\citeyearads}[3][][]%
{\href{http://adsabs.harvard.edu/abs/#3}
{\def\hyper@linkstart##1##2{}%
\let\hyper@linkend\@empty\citeyear[#1][#2]{#3}}}
\makeatother

\begin{document}

\title{Identification of coronal heating events in 3D simulations}


\author{Charalambos Kanella
          \and
          Boris V. Gudiksen}

\institute{Institute for Theoretical Astrophysics, University of Oslo, P.O. Box 1029 Blindern, N-0315 Oslo, Norway\\
\email{charalambos.kanella@astro.uio.no}}

\abstract
{The  \emph{solar coronal heating problem} has been an open question in the science community since 1939. One of the proposed models for the transport and release of mechanical energy generated in the sub-phorospheric layers and photosphere is the magnetic reconnection model that incorporates Ohmic heating which releases a part of the energy stored in the magnetic field. In this model  many unresolved flaring events occur in the solar corona, releasing enough energy to heat the corona.}
{The problem with the verification and quantification of this model is that we cannot resolve small scale events due to limitations of the current observational instrumentation. Flaring events have scaling behavior extending from large X-class flares down to the so far unobserved nanoflares. Histograms of observable characteristics of flares, show powerlaw behavior, for both energy release rate, size and total energy. Depending on the powerlaw index of the energy release, nanoflares might be an important candidate for coronal heating; we seek to find that index.}
{In this paper we employ a numerical 3D-MHD simulation produce by the numerical code {\it{Bifrost}} , which enable us to look into smaller structures, and a new technique  to identify the 3D heating events at a specific instant. The quantity we explore is the Joule heating, a term calculated directly by the code, which is explicitly correlated with the magnetic reconnection because it depends on the curl of the magnetic field.}
{We are able to identify 4136 events in a volume $24 \times 24 \times 9.5 \ \textrm{Mm}^3$ (i.e. $768 \times 786 \times 331$ grid cells) of a specific snapshot. We find  a powerlaw slope of  the released energy per second equal to $\alpha_P=1.5 \pm 0.02$, and two powerlaw slopes of the identified volume equal to $\alpha_V=1.53 \pm 0.03$ and  $\alpha_V=2.53 \pm 0.22$ . The identified energy events do not represent all the released energy, but of the identified events, the total energy of the largest events dominate the energy release. Most of the energy release happens in the lower corona, while heating drops with height. We find that with a specific identification method that large events can be resolved into smaller ones, but at the expense of the total identified energy releases. The energy release which cannot be identified as an event favours a low energy release mechanism. }
{This is the first  step to quantitatively identify magnetic reconnection sites and measure the energy released by current sheet formation.}
   
   \maketitle

\section{Introduction}

The solar corona is counter intuitively much hotter than the solar photosphere. Energy cannot be directly transported from the photosphere to the corona through heat conduction or mass motions making the so-called ``coronal heating problem" into two problems: An energy transport problem, and a dissipation problem. It has been shown that sound waves are not able to carry energy into the corona (\citetads{2002ApJ...572..626C}; \citetads{2007PASJ...59S.663C}), leaving the magnetic field as the main ingredient in the transport problem. 

The tremendously large mechanical energy generated by the motions of the photosphere and the underlying convective layers can heat the solar corona if energy can be transported and released there \citep{Goldbook}. The magnetic field, anchored in the photosphere, extends through the solar atmosphere, enabling the mechanical energy to propagate via Poynting flux towards the corona.
If the energy initially has no release mechanism, the energy is stored increasingly in the magnetic field until an instability occurs and suddenly a fraction of the stored energy is released. 

Since the ceaseless motion of the anchored magnetic field shuffles the magnetic structures, magnetic gradients can increase and magnetic reconnection triggered, creating the instability necessary to dissipate the stored energy (\citetads{1983ApJ...264..635P}; \citetads{1983ApJ...264..642P}). This is the mechanism of flares and due to the large magnetic field gradients, current sheets are formed at the reconnection sites. For a given distribution of magnetic polarities in the photosphere, the potential current-free field in the atmosphere is the lowest possible energy state of the field. A field configuration with the same photospheric polarity distribution, but a higher energy, have the energy stored in electric currents distributed throughout the field. As a consequence, a part of the stored magnetic energy is released via Ohmic (Joule) heating \citepads{Low1990}, leaving behind a newly formed magnetic configuration simpler than before \citepads{1996JGR...10113445G}. 

According to \citetads{1996RSPTA.354.2951P}; \citetads{2002A&ARv..10..313P} and \citetads{2009PhPl...16l2101P} there are four types of reconnection each with a dissipation site with a different shape. First, there is the so-called ``spine reconnection" in which the current and therefore the dissipation site is located along a field line that passes infinitely close to a null point, the spine. Secondly,  there is the so called ``fan reconnection", in which the dissipation site is located at the fan surface (a set of field lines that leaves or approaches the nul point forming a plane incorporating the null point. Separator reconnection in which the dissipation sites lay along a separator that is separating two or more regions where the topology of the magnetic field is different. The final type of reconnection is in a  ``quasi-separatrix layer'' (QSL). Here no null points or seperators need to be present, but are regions where the magnetic field has large gradients. Reconnection in QSLs  is related to the slip-running reconnection as shown by \citet{Janvier2013} in numerical simulations. Which type of reconnection occurs depends on the nature of the footpoint motions and the structure of the magnetic field. 

According to \citetads{1972ApJ...174..499P} and observations \citepads{1988ApJ...330..474P}, the \emph{nanoflare} heating model can heat the solar corona, if a very large number of these events occurs endlessly in order to satisfy the energy requirements of the corona to sustain the high temperature. Despite the progress of the instrumentation employed to observe the sun and especially the flaring events, it is still impossible to observe nanoflares.  According to observations, the frequency of energy release from flaring events is distributed as a powerlaw function, $N(E) \propto E^{-\alpha}$, where $\alpha$ is the power index and $N(E)$  the number of events  in the energy range $E$ and $E + \delta E$. What makes  the power index important is the fact that if the index is larger than  two, then nanoflares have a larger contribution to the energy budget than large flares \citepads{1991SoPh..133..357H}, thus making the nanoflare model a powerful candidate for coronal heating.

Numerous theoretical investigations into how the size distribution (i.e. powerlaw distribution) of flares are formed (\citetads{1991ApJ...380L..89L}; \citetads{1996ApJ...469L.135G}; \citetads{2008ApJ...682..654M}; \citetads{2011SSRv..159..263H}; \citetads{2014SSRv..tmp...29A} suggest that flares are manifestations of loss of equilibrium in a system that evolves naturally towards a non-equilibrium state.  Flares appear when a critical point is passed and  the system relaxes by releasing a fraction of the stored energy and thus, self-organisation is achieved and the system reaches a less stressed state. This picture was conceived first by \citetads{1988PhRvA..38..364B} for avalanches in sandpiles and used in solar physics by \citetads{1991SoPh..133..357H}, in which the powerlaw size distributions is a result of the stochastic energy release. In other words, it is impossible to predict how much energy the system will release when loss of equilibrium occurs, but we can expect that small events are more likely to occur than larger ones, and the maximum energy release that could be released is the total excess of energy stored in the magnetic field.

It is extremely difficult to specify the powerlaw index of the frequency distribution because of the observational biases. Observations struggle with the determination of structures along the line of sight and the noise from background and foreground sources. For instance, observations of flaring events suggest that the spectral index of the thermal energies for M- and X-class GOES class flares is $\alpha_{th} =1.66 \pm 0.13$ \citepads{2013ApJ...776..132A}. However, the determination of thermal energies requires the knowledge of the volume occupied by the flaring events \citepads{2002ApJ...568..413B}. Since, the projected area is the only geometrical aspect extracted from observations, scaling laws must be used in order to make assumption for the third dimension (e.g. \citetads{2002ApJ...568..413B}; \citetads{2013ApJ...776..132A}). In addition, as revealed by \citetads{2009ApJ...698.1893M}, the projected area of flares and thus, the assumed volume depends on the projection (e.g Fig. 3 in the same reference). For instance, different spectral lines target different depths in the solar atmosphere and therefore, different flare areas are observed. Also, the temperature and emission measure needed to calculate thermal energy can be found through the fitting of an isothermal model spectrum to either the flare spectrum or the ratio of images in multiple wavelengths. Conclusively, the ambiguities of the models fitted in combination with the bias on the observational parameters result in large uncertainties.  

A broad range of powerlaw slopes has been derived for different size distributions (see table 4 in \citetads{2014SSRv..tmp...29A}). Different slopes of the frequency of occurrence of solar flares in different passbands have been extracted, e.g. $\alpha = 1.2 - 2.1$ for peak fluxes and  $\alpha = 1.4 - 2.6$ for powerlaw slope of total fluence in EUV, UV and H$\alpha$.
These results point out a high dependence of the statistical analysis on many parameters such as: the method used to detect and select flaring events, the event definition, the synchronicity of the images in different spectral lines or filters, the goodness of sampling at the lower (due to finite resolution) and higher ends (due to incomplete sampling, which causes truncation effects), the number of events, the identification of the duration of a flaring event, the fitting methods and the error bars used in fitting, and finaly, the correct subtraction of background heating and noise. A combination of all these criteria are found to affect the powerlaw index (\citetads{2000ApJ...529..554P}; \citetads{2002ApJ...568..413B};  \citetads{2015ApJ...814...19A}). As stated also by \citetads{2011SSRv..159..263H}, the large range of powerlaws in different studies from various researchers (see Fig. 2 in \citetads{2011SSRv..159..263H}) depends on the methodology and instrumentation used in different periods during the solar cycle.

The last confusing factor is that the the powerlaw index quoted are often for very different things, such as size, max temperature, fluence in different filters etc which is also done above. 

Observations have so far not been able to decide the importance of flares in coronal heating, so in this paper we will use a different method. 3D MHD numerical models together with the correct identification algorithm can detect flaring events and extract them from the simulation data and flare parameters can be derived. Because of the completeness of the data in the simulation we can derive the following parameters for each flare (heating event): volume, energy release, maximum energy release rate and height of formation. The correctness of these parameters of course rely on the numerical code being able to reproduce a realistic model of the solar atmosphere. This paper will not discuss the ability of the numerical code to do so \citepads[see][for such an investigation]{2009ApJ...694L.128L}, but merely describe our post processing methods and results of those methods. 

The remainder of this paper is organised in the following way:  in Sect. \ref{subsec: Bifrost} we briefly describe the {\it{Bifrost}} code \citep{Gudiksen2011} used to simulate the sun from the convection zone up to the corona. In Sect. \ref{subsec:Identification_method}, we explain the method used to identify heating events, while in Sect. \ref{sec:Results}, the results of our investigation along with the statistical analysis (i.e in \ref{subsec:statistical_analysis}) are discussed. Finally, in Sect. \ref{sec:conclusions} we sum up the main points of the current work and discuss in detail what we find.
	
\section{Method} \label{sec:Method}

\subsection{Bifrost Simulation} \label{subsec: Bifrost}


The {\it{Bifrost}} code \citep{Gudiksen2011} is a massively parallel code, which can simulate a stellar atmosphere environment from the convection zone up to the corona. It can include numerous special physics and boundary conditions so to describe the stellar atmosphere in a more detailed manner. It solves a closed set of 3D MHD partial differential equations together with equations that describes radiative transport and thermal conduction. The system of equations is solved on a Cartesian grid using 6th order differential operators, 5th order interpolation operators along with a 3rd order method for variable time-step.

The code includes different processes occurring in the convection zone, photosphere, chromosphere, transition region and corona. In the convection zone, the code solves the full radiative transfer including scattering. The code treats the radiative transfer in the corona, as optically thin. The region of the chromosphere, where the atmosphere is optically thin for the continuum, but optically thick for a number of strong spectral lines, the radiative losses are calculated from the procedure derived by \citet{CarlssonLeenaarts2012}. 

Both the radiative and conductive processes are described via the equation of internal energy which has the following form: 
\begin{equation} \label{eq:energy}
\frac{\partial e}{\partial t} +  \vec{\nabla} \cdot e\vec{u} = Q_{\text{c}} - \Lambda - P \vec{\nabla} \cdot \vec{u} + Q_{ \text{J}} + Q_{ \text{Vi}} 
\end{equation}
where $e$ is the internal energy per unit volume, $\vec{u}$ the velocity vector, $P$ the gas pressure, $Q_{\text{c}}$ the heating/cooling derived via the Spitzer thermal conduction along the magnetic field \citep{Spitzerbook}. $ Q_{ \text{J}}$ represents the Joule heating, $Q_{ \text{Vi}}$ is the viscous heating and $\Lambda$ the cooling or heating produced by the emission and absorption of radiation. 

In the current work, we simulate the region from the solar convective zone up to the corona. The simulated volume starts 2.5 Mm below the photosphere and extends 14.3 Mm above the photosphere into the corona.  We use periodic boundary condition in the horizontal x--y plane; in the vertical z--direction, the upper boundary is open,  while the lower boundary is open, but remaining in hydrostatic equilibrium enabling convective flows to enter and leave the system. By controlling the entropy of the material flowing in through the bottom boundary, the effective temperature of the photosphere is  kept roughly constant at $5.780 \times 10^3 $ K.  The simulation box has a volume equal to $24 \times 24 \times 16.8 $ Mm$^3$, which is resolved by  $768 \times 768 \times 768$ cells. Therefore, the horizontal grid spacing $dx = dy = 31.25$ km, whereas the vertical grid spacing varies to resolve the magnetic field, temperature and pressure scale heights. Hence, the vertical spacing ($dz$) is approximately equal to 26.09 km in the photosphere, chromosphere and transition region, while increases up to 165 km at the upper boundary in the corona. This simulation incorporates  two strong magnetic regions of opposite polarity, which are connected with a magnetic structure with loop-like shape. The magnetic field is initially set vertically at the bottom boundary and extrapolated to the whole atmosphere assuming potential field, while a horizontal 100 Gauss field is fed continuously in at the lower boundary producing random salt and pepper magnetic structures in the photosphere. Further details of the simulation setup can be found in \citet{Carlsson2016} which explains a similar setup, except the one describe in \citet{Carlsson2016} also includes the effects of non-equilibrium ionisation of hydrogen.

To quantitatively study the effects of magnetic reconnection, we choose to analyse the joule heating term which is calculated directly in {\it{Bifrost}} through Ohms law, using a non-constant electric resistivity. The electric resistivity is kept as low as possible, while still stabilising the code. That means that the resistivity is not uniform in the computational box and it is everywhere larger than the microscopic resistivity. The resistivity in the highest current regions is consequently larger than everywhere else, but the sun also seems to have a method to increase the resistivity locally, otherwise fast reconnection would not be possible (\citetads{1986mrt..conf...19B}; \citetads{1989JGR....94.8805S}). How the sun is able to increase the electrical resistivity locally is not known, but several methods are possible, such as instabilities in the central current sheet leading to small scale turbulence. How the resistivity behaves is unknown, except that it does increase in reconnection sites in the solar atmosphere. {\it{Bifrost}} therefore uses the most conservative value of the electric resistivity available, which is the value of the resistivity which keeps the code stable. 

\subsection{Identification Method}\label{subsec:Identification_method}

We choose a region of interest (ROI), which starts from the lower corona, 3.28 Mm above the photosphere, where the temperature is on average equal to 1 MK, and extends up to the top of the simulation box. The volume of interest is resolved by $768 \times 768 \times 331$ grid cells, spanning $24 \times 24 \times 9.5$ Mm$^3$. 

Identifying locations with current sheets seems at first simple. Locating 3D volumes with high Joule heating should be easy enough, but it turns out not to be so simple. The problems we encounter is that the current sheets in 3D is generally not 2D flat structures as the cartoon like pictures of 2D reconnection would suggest, but much more complex. Often the ``background'' current level is higher in places with many current sheets, so it is not easy to separate one current sheet from another. That is in some ways similar to the problems experienced by observers, where the background is giving large problems for the interpretations. The method we have identified as the best is a powerful numerical tool, named ``ImageJ'', used in medical imaging and bio-informatics to perform multi-dimensional image analysis \citep{Ollion2013, Gul-Mohammed2014}. ImageJ is an open source tool in which users contribute by creating plugins for different purposes, making them available to everyone. In our case, we use the plugin ``3D iterative thresholding" (also known as AGITA: Adaptive Generic Iterative Thresholding Algorithm, \citepads{Gul-Mohammed2014}), which first, detects features for multiple thresholds, and then it tries to build lineage between the detected features from one threshold to the next one, and finally choose the features that fall within the pre-specified criteria. Therefore, different features can be segmented at different thresholds. The thresholds tested here can be chosen based on three value parameters. The first parameter is the ``step" parameter: the threshold increases by the specific input value (step) and features are identified between the previous threshold and the current one. The second parameter is ``Kmeans", in which a histogram is analysed and clustered into the specific number of pre-specified number of classes. The last parameter is the ``volume" parameter, for which the method tries to find a constant number of pixels between two thresholds for different threshold values in each step. In addition, for each lineage (same features segmented at different thresholds) the plugin allows various criteria to pick the best identified feature. For instance, the ``elongation" criterion chooses the most rounded features, the ``volume" criterion chooses the largest features, and the ``MSER" criterion chooses features with the smallest variation of number of pixels for different thresholds (i.e. the feature with the smallest number of pixels).

In our case, we first load our data, and then scale them from a 32-bit 3D image  to 16-bit; instead of having single precision floats, we now have integers that scale from 0 to 65535. The new image is scaled dynamically according to the upper and lower limits of the energy rate density. We then, use the iterative thresholding method using the ``step" parameter equal to 50. The chosen step-parameter is based on a combination of making sure that the produced features do not change too much between successive thresholds, while still not being too computational expensive to traverse the range from 0 to 65535. We found a step size of 50 to be sufficient. 

We also choose a minimum and maximum number of cells in an feature equal to 125 and 100000 respectively; the minimum value is 125 because {\it{Bifrost}} uses 6th order differential and 5th order interpolation operators in a stagger grid where the computational stencil is 6 grid points. Choosing 5 grid cells in each direction is a very conservative choice, because the the 6th order operators should be able to resolve a peak with just 3 points. We also, choose the volume criterion so that the method chooses the largest possible features, because we want to make sure that the method do not split features that are truly one feature, thereby changing the powerlaw distributions that are the goal of this investigation. The plugin then labels each feature with a specific number and we store each newly generated 3D image which contains the labels.

The method resolves approximately $12\%$ of the total Joule heating. The unresolved energy is considered to be a combination of unresolved features and background heating that is not correlated with impulsive events and effects of the imperfect method to separate features. 

The background heating varies almost as much as its amplitude, and we have so far not been able to produce a method that would not muddle the final statistics.

General characteristics can be derived after there identification of heating events. We calculate the spectral index $\alpha$ of the powerlaw distribution of the energy rate density and volume of the resolved objects, using the method proposed by \citetads{2015ApJ...814...19A}. This method should counter the problems of: the physical threshold of an instability may change for any physical reason, incomplete sampling, and contamination by event-unrelated background affecting the frequency of occurrence of small scale events.The largest events will naturally be missing due to the limited volume and energy density of the ROI. The powerlaw fitting consequently has to take these limitations into account, and more free parameters then need to be included in the fit:

\begin{equation}\label{eq:power_law}
N(x) = n_0 * (x+x_0)^{-\alpha}
\end{equation}
where $x$ denotes the specific quantity, $n_0$ is a constant, which depends on the total number of features ($n_{f}$) in a range bounded from the maximum ($x_2$) and minimum ($x_1$) values

\begin{equation}
n_0 = n_{f}(1-\alpha)[x_2^{1-\alpha} - x_1^{1-\alpha}]^{-1}
\end{equation}
where $n_{f} = \int^{x_2}_{x_1} N(x) dx$ is the result of normalisation. Thus, we try to fit $n_0$, $\alpha$ and $x_0$ of equation \ref{eq:power_law} via $\chi^2$ minimisation. The specific powerlaw fitting technique reduces the number of data points needed to fit the powerlaw, but in our case the difference between the identified events and the reduced ones is insignificant.

\section{Results} \label{sec:Results}

The method mentioned in the previous section represents an effort to identify discrete heating events. We try to separate those events from other sources of heating that are not included in the Joule heating term. 

Our method identifies initially large features, which are clusters of events and each comprised of more than $10^4$ grid-cells. They are mostly located at the bottom of the corona, where most of the current sheets are formed, a result similar to what was observed by \citetads{2005ApJ...618.1020G}. Features that have large horizontal cross section at the bottom of the corona leads to large volume when we perform the clustering method of \citetads{1976PhRvB..14.3438H}. Looking at a 3D rendering of the large events and plotting the energy release along random lines throughout their volumes, it is clear that very large events are collections of many small events closely packed in space. Our method is clearly not able to identify features which are this closely packed. 

The volume of the events is  $4\%$ of the total coronal volume and the total energy contained in all events is only 12\% of the total Joule heating in the corona. The remaining 88$ \%$ of joule heating is a combination of unsatisfied threshold criteria which either do not satisfy the minimum resolution criterion or the method is unable to find the full extension of an event. The unresolved features with a volume less than 125 grid cells account for less than $0.1\%$ of the total joule energy term in ROI, so the unresolved features are not a significant source of error.

\begin{figure}
\centering
\includegraphics[width=\hsize]{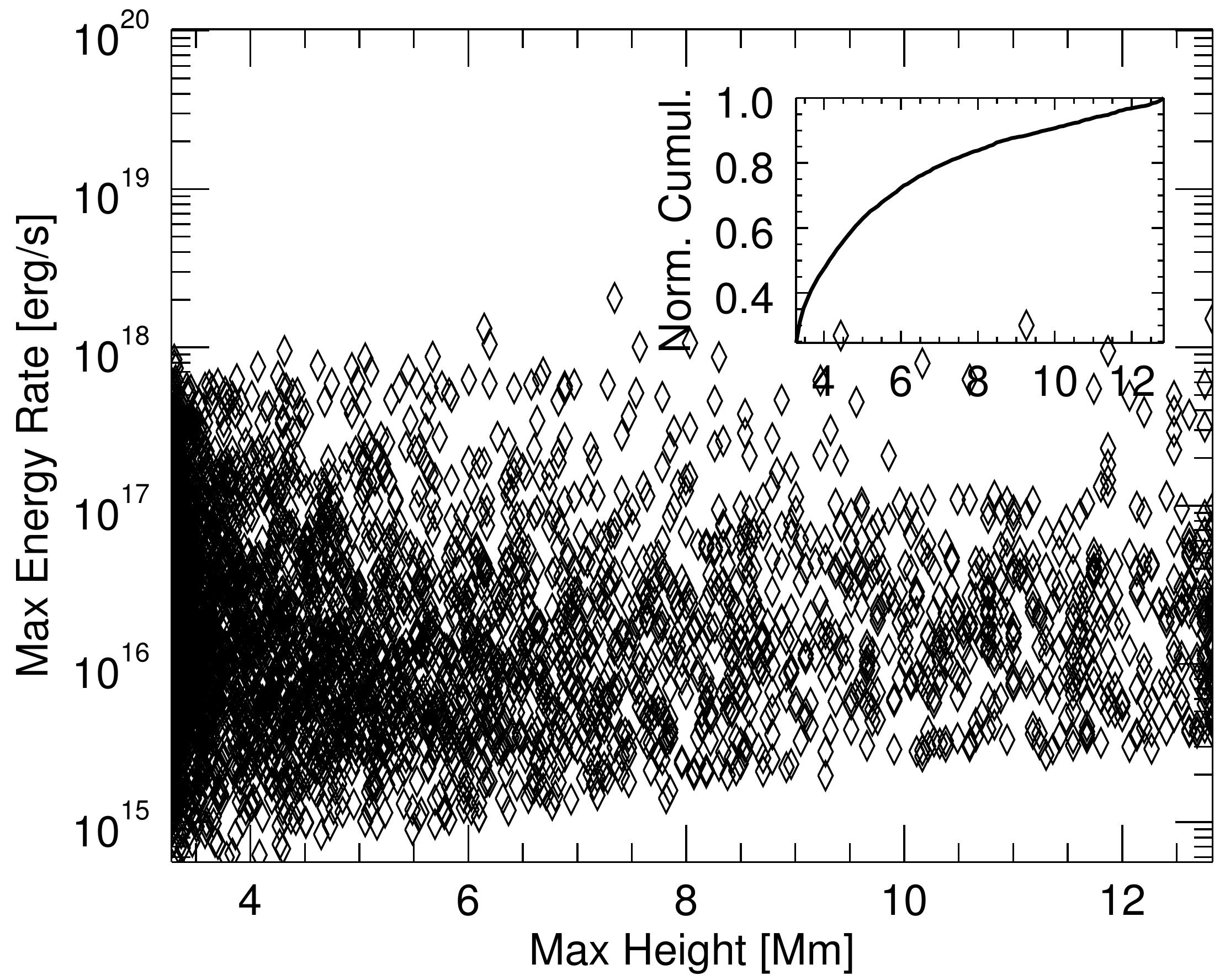}
\caption{Plot of height of max energy rate for each identified feature with respect to the value of the maximum energy rate plotted in logarithmic scale; this height may be correlated with the height of event initiation. It is clear that the activity in the lower corona is higher than in the rest region. From the normalized cumulative plot we observe also the same result: the biggest percentage of data points lays at the lower corona. \label{fig:en_vs_height}}
\end{figure}

Most of the heating events are located at the bottom of the corona, a fact that can be deduced from the density of features in Fig. \ref{fig:en_vs_height}. Both illustrations in Fig. \ref{fig:3d_image} show that the heating events extend upward or hover in the corona. The analysis also shows that they usually have three shapes: fan-like, spine-like, and loop-like shapes.  The same shapes have also been observed in simulations by \citetads{2015ApJ...811..106H} and described extensively by \citetads{2002A&ARv..10..313P}, where currents form around the fan-, spine- and loop-like structures of the magnetic field during reconnection.

\section{Statistical Analysis}\label{subsec:statistical_analysis}

\begin{figure}
\begin{subfigure}{0.5\textwidth}
\centering
\includegraphics[width=\hsize]{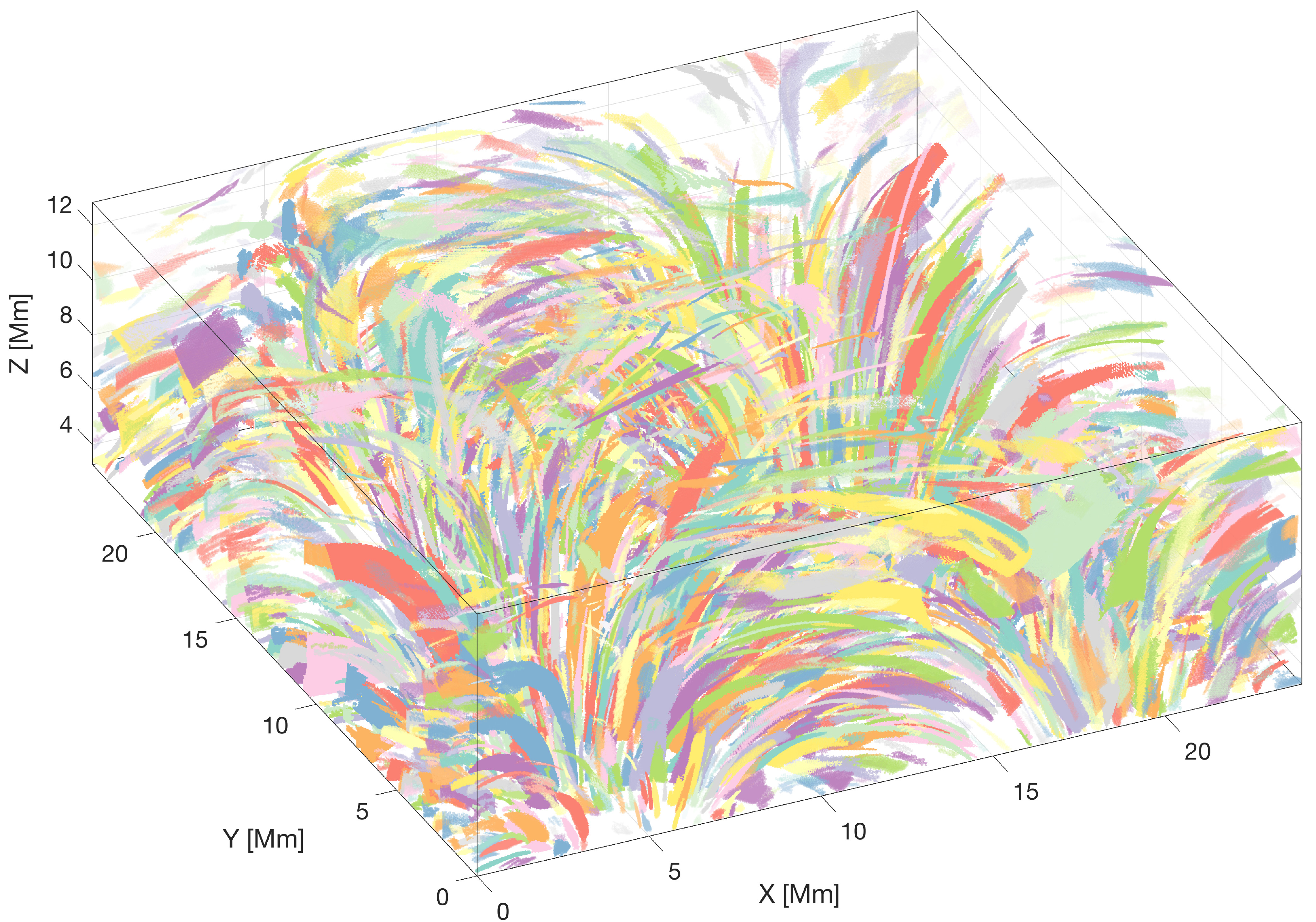}
\end{subfigure}
\begin{subfigure}{0.5\textwidth}
\centering
\includegraphics[width=\hsize]{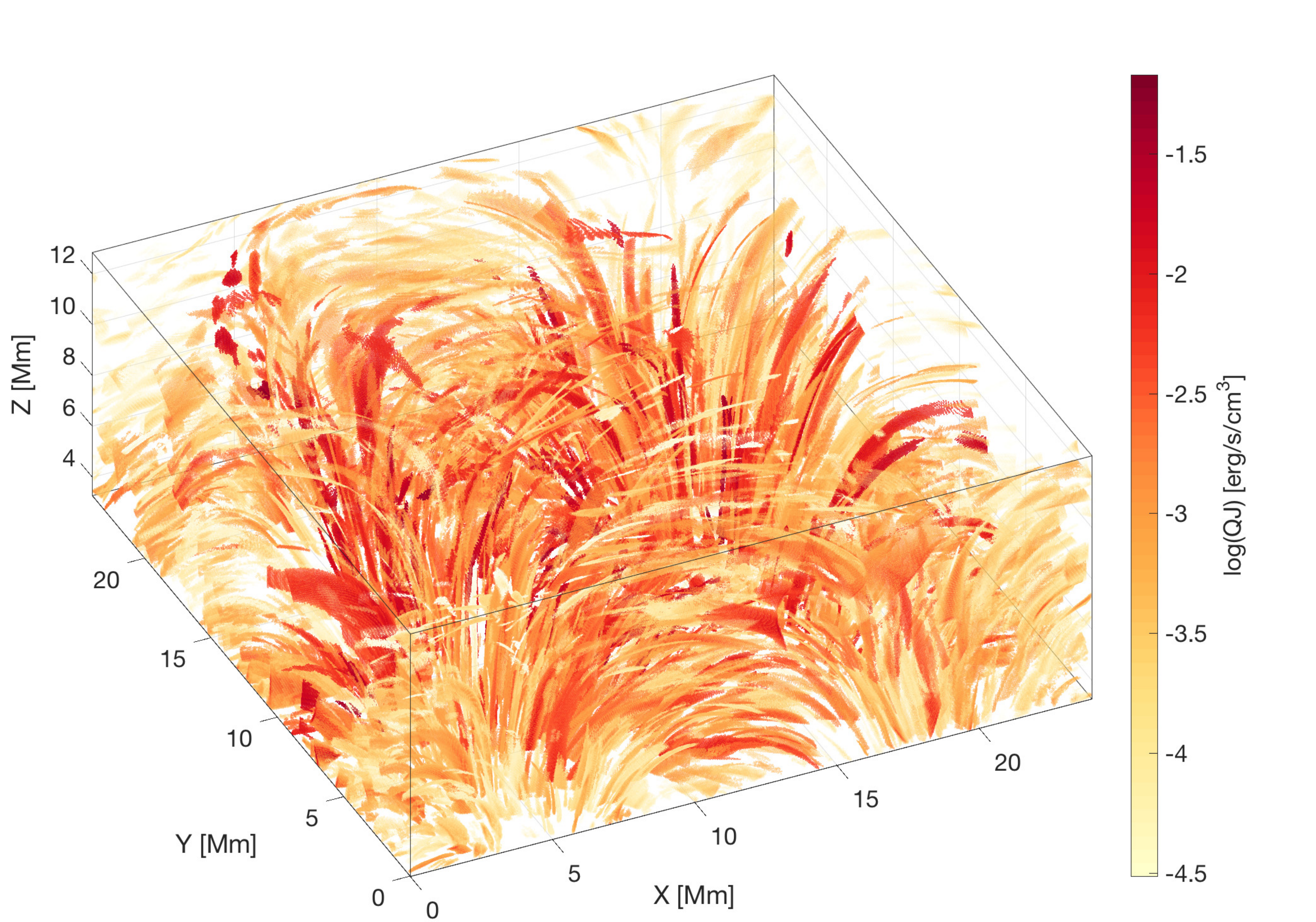}
\end{subfigure}
\caption{Top: 3-D rendering of 4136 identified features of the joule heating term, in which each color represents different feature. Bottom: indentified features of joule heating in logarithmic scale. The energy output from the resolved features is 12$\%$ of the joule heating output, while the rest 88$\%$ counts as background heating, numerical noise and unresolved features.  \label{fig:3d_image}}
\end{figure}

In Fig. \ref{fig:3d_image} we are seeing the result of the heating events imprinted at a random moment in their life, and in the following we do statistics on these events. 

\begin{figure}
\centering
\includegraphics[width=\hsize]{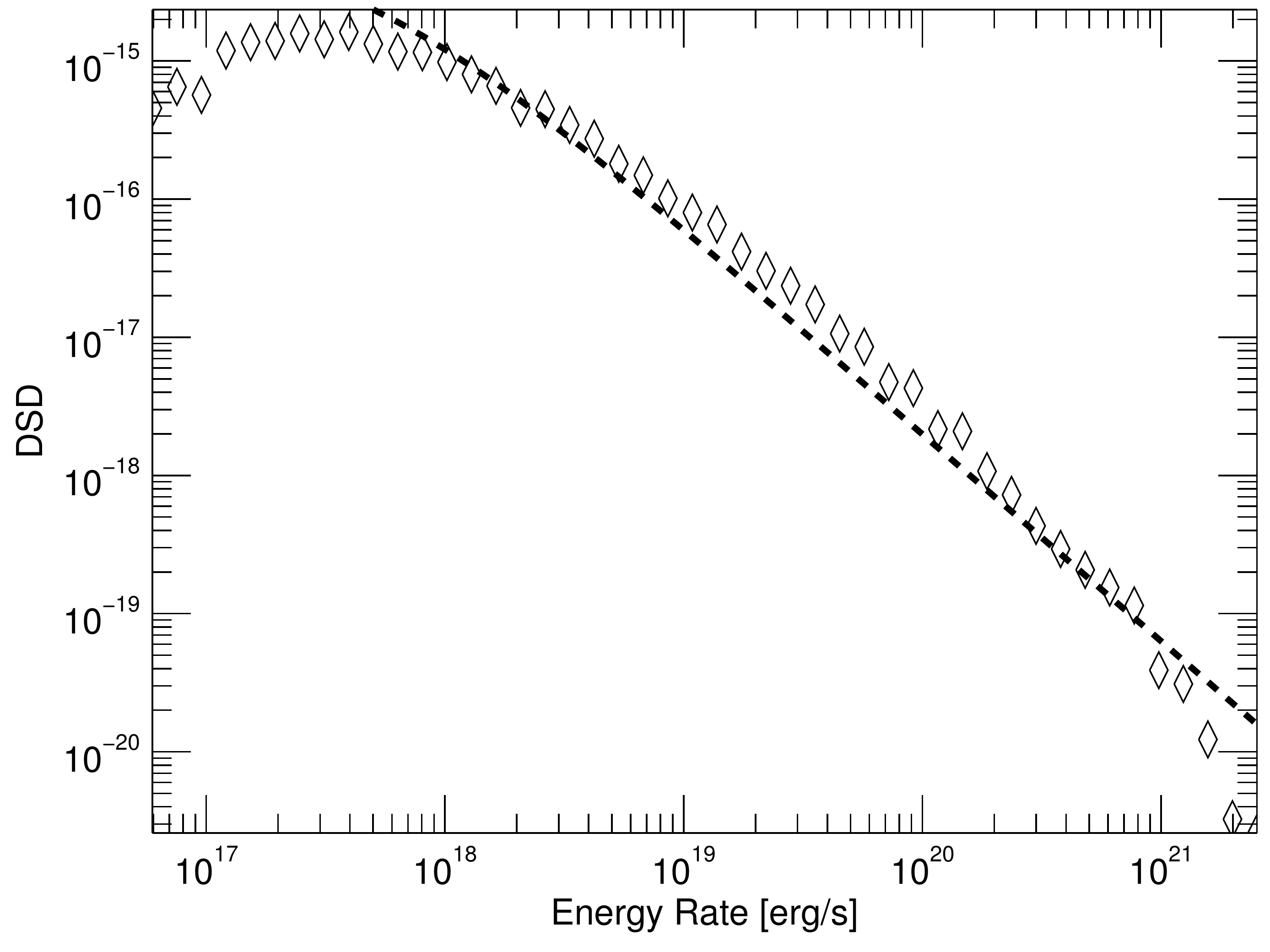}
\caption{Plot of differential size distribution  of the identified features' energy rate in logarithmic scale representing $N=4136$ identified features (diamonds). A powerlaw index $\alpha=1.50 \ \pm 0.02$ is found, having goodness-of-fit $\chi^2=14.28$; the error is calculated as $\alpha/\sqrt{NF}$, where the number of features of the fitted function is $NF=3818$ (dashed line), excluding data with energy rate less than background value $x_0=4\times 10^{17} \ \textrm{erg/s}$. \label{fig:hist_ene}}
\end{figure}

In Fig. \ref{fig:hist_ene}, we plot the differential size distribution of released energy rate  over logarithmic bin. The differential size distribution is defined as the number of features with sizes within a certain bin, divided by the bin size. To calculate the number of bins we use the formula derived by\citetads{2015ApJ...814...19A}. The formula has the following form:

\begin{equation}
n_{bin}= 10\, \log_{10}\left(\frac{x_{max}}{x_{min}}\right)
\end{equation}

where $x_{max}$ and $x_{min}$ is the maximum and minimum value in the data set. We notice that the distribution of the released energy rate follows a powerlaw function, which has the following form:

\begin{equation}
y=n_0 x^{-\alpha} 
\end{equation}

where the powerlaw index has value $\alpha=1.5 \pm 0.04$. The error is calculated as $\alpha/\sqrt{NF}$, where $NF=1368$ is the number of events and the powerlaw function is fitted within a range of released energy rate spanning from $2 \times 10^{18} \ \textrm{erg/s}$ to $2 \times 10^{22} \ \textrm{erg/s}$.

\begin{figure}
\centering
\includegraphics[width=\hsize]{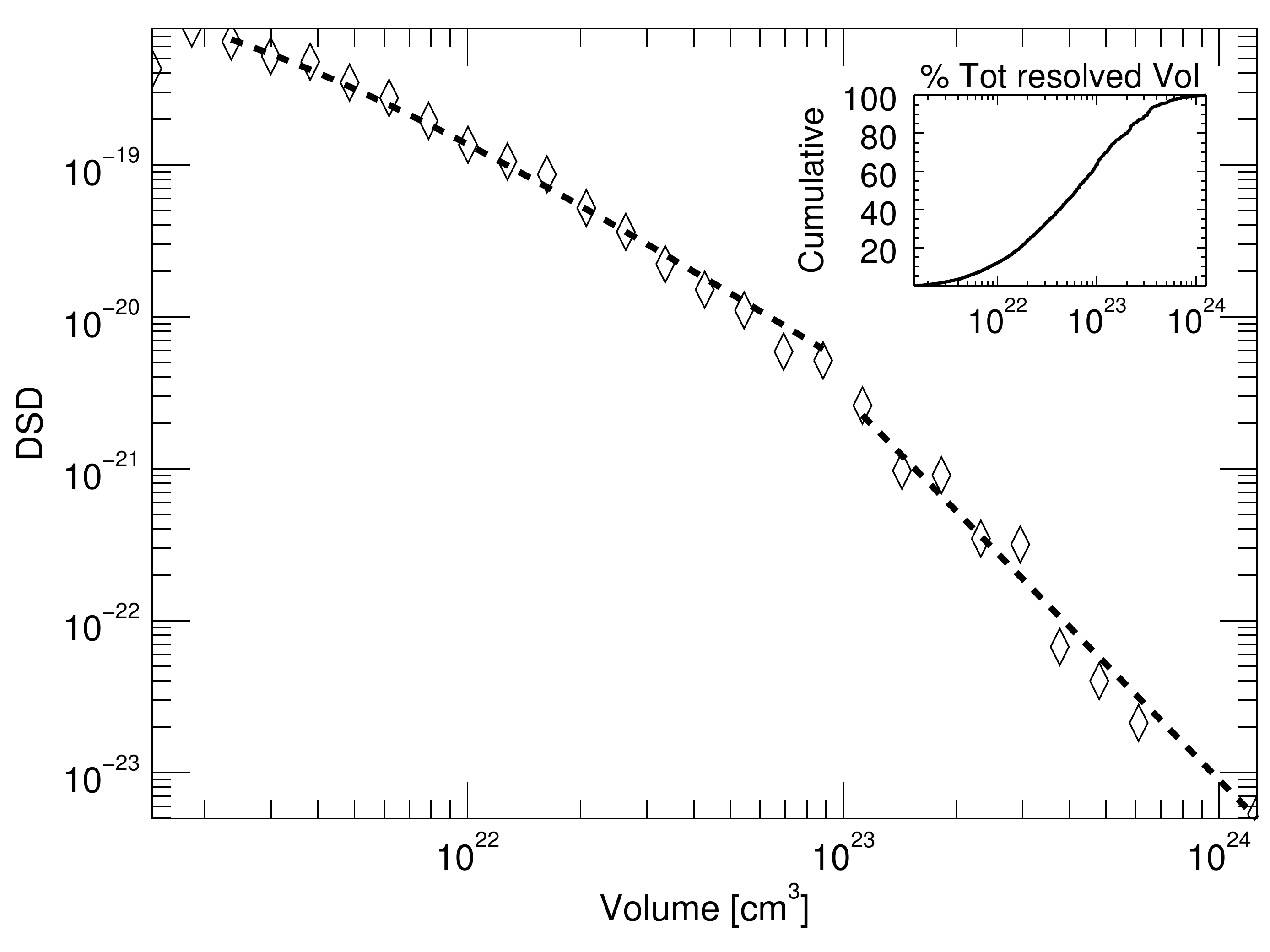}
\caption{Plot of differential size distribution of the identified features' volume in logarithmic scale. Two powerlaws are fitted to the data, which have slopes equal to $\alpha = 1.53 \pm 0.03$ with a goodness-of-fit $\chi^2 =3.43$ and $\alpha = 2.53 \pm 0.22$ with a goodness-of-fit $\chi^2 =1.28$ respectively. The first slope is fitted over 3542 data points in the range between $1.85 \times 10^{21} $ and $10^{23} \ \textrm{cm}^3$, whereas the second over 137 data points in the range between $10^{22} $ and $10^{24} \ \textrm{cm}^3$. In the right corner the cumulative of the resolved volume is plotted, i.e. $N(< x)=f(x)$. \label{fig:hist_vol}}
\end{figure}

The volume of events is also important. In Fig. \ref{fig:hist_vol}, we plot the differential size distribution of volume, along with the fitted powerlaw function. What we find is two spectral indices $\alpha_V=1.53 \pm 0.03$  within a range between $10^{21}$ and $10^{23} \ \textrm{cm}^3$ and $\alpha_V=2.53 \pm 0.22$  within a range between $10^{23}$ and $2 \times10^{24} \ \textrm{cm}^3$ . 
To find which scale of events' volume is the most important in terms of space filling plot the cumulative size distribution of this size in Fig. \ref{fig:hist_vol}. It seems that large scale events, i.e. structures larger than $10^{23} \ \textrm{cm}^3$, which correspond to 137 events occupy 40\% of the resolved volume in the ROI. 

\begin{figure}
\centering
\includegraphics[width=\hsize]{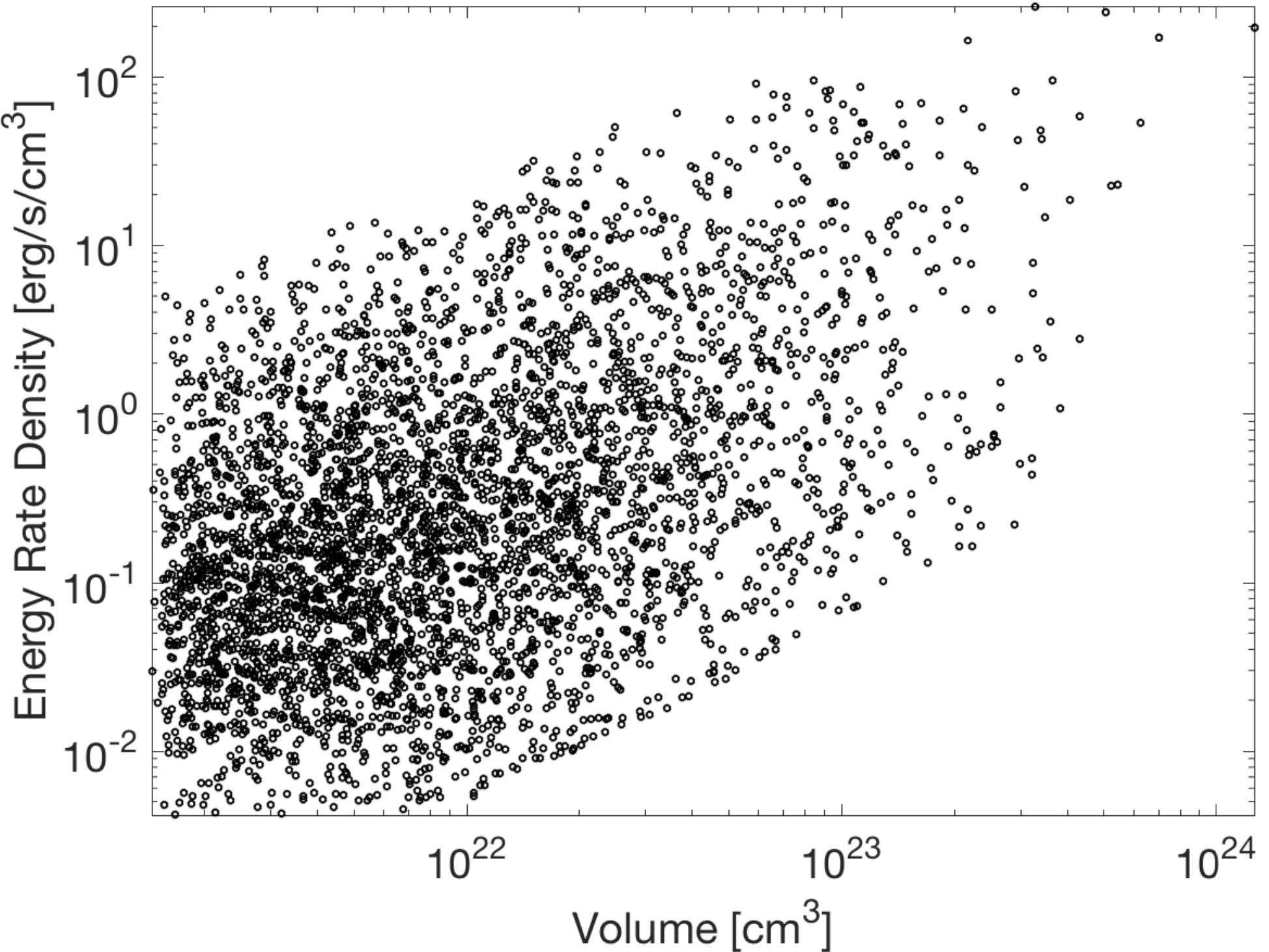}
\caption{Plot of energy release rate over volume versus volume of identified features in logarithmic scale.  \label{fig:vol_powerdens}}
\end{figure}

We also check for any correlation between volume and energy release per volume. Figure \ref{fig:vol_powerdens} depicts the two quantities plotted against each other for all identified events.  The two quantities seem not to have any clear correlation. Features of any volume span more than 3 orders of magnitude, and features with any energy rate density spans two and half orders of magnitude in volume. 

To further test the connection between volume and the energy release rate per volume, we perform Spearman's rank and Pearson linear correlation. The rank correlation showed a merely bad global correlation i.e. $\rho=0.4$ (this correlation varies between 0 and 1, smaller values indicate good rank correlation), whereas there is  very weak linear correlation (i.e. $\rho=0.53$) between the two quantities (Pearson correlation varies between -1 and 1). 

Figure \ref{fig:en_vs_height} shows the maximum energy release rate of each event as a function of the height that the specific maximum energy release is located. 

If we assume that  the location where an event is triggered, 
then the plot suggests possible height of instability triggering. 

From the normalised cumulative plot, and from the plot itself, we observe that the lower corona contains not only most of the heating events but also the most energetic ones. Almost $40\%$ of the total number of events are located between 3 and 4 Mm above the photosphere.  The number of events in higher heights up to 14 Mm above the photosphere is distributed almost evenly while the maximum energy release rate drops with height. Therefore, the heating in the lower corona is larger than in the upper layers, a trend also observed by \citetads{2002ApJ...572L.113G}. 

\section{Disussion and Conclusions}\label{sec:conclusions}

This paper discusses the implementation of a multi-thresholding technique implemented on 3D MHD simulation aiming to identify three dimensional joule heating events at a randomly selected snapshot from a numerical simulation of the solar corona. 

Using the {\it{Bifrost}} code, we simulate the solar environment enabling us  to identify events in the modelled solar corona with high resolution. Those events release power which spans almost 4 orders of magnitude starting from $10^{17} \ \textrm{erg/s}$ , and have volume, which spans 3 orders of magnitude starting from $10^{21} \ \textrm{cm}^3$. The events follow a powerlaw over many orders of magnitude  just like many other self-organised critical systems, suggesting that the formation of these structures share the same physical mechanism that scale in the energy and volume regimes. 

The outcome is the result of the stochastic nature of magnetic reconnections ability to release energy stored in the magnetic field, when it reaches a threshold. The stochastic nature originates from the fact that magnetic reconnection triggers an instability in which a random fraction of the energy stored in the magnetic field is released. In some cases observations shows that this system appears to have memory of previous energy releases as magnetic reconnection events are sometimes observed to happen at the same location within a short amount of time, i.e. homologous flares \citepads{1967SoPh....2..316F}. This is in our opinion caused by the stochastic nature of the total energy release. If the energy released is relatively small compared to the surplus energy stored in the magnetic field at a specific location, then the fact that energy is released might produce conditions where only a small increase in the stored energy can lead to yet another energy release.

We find that there is no global linear relation between energy release and volume, and the Spearman's rank correlation shows a merely bad correlation.  

Generally, the heating is mostly concentrated at the bottom of the corona and gradually drops with height because the magnetic field magnitude also drops with height; a fact that was pointed out and explained by \citetads{2002ApJ...572L.113G}. A consequence of the distribution is that energetic events are more likely to be generated in the lower corona.
 
The differential size distribution of released energy rate and volume follow a powerlaw distribution. What we find are slopes that favour the release of energy in large events. 

As illustrated in Fig. \ref{fig:3d_image}, the method can resolve large structures into smaller and identify where most of the heating occurs locally even though some of the energy is no longer in identified events. The unresolved energy might be attributed to two categories of reasons: technical and physical reasons. In the first category, unsatisfied thresholding criteria, unresolved features or numerical heating due to noise have an impact on identifying real heating events. The unresolved energy could be due to heating from other sources, such as MHD waves that distort the magnetic field, or remnants of currents sheets after energetic events that burn slowly \citepads{2014ApJ...788...60J}. Finding dissipating MHD wave modes in this snapshot is outside the scope of this work. 

Our method cannot identify all the energy released as events. If the large part of the total Joule heating which is not identified as events in this work, actually is small events, then they would be added to the lowe energy tail of our powerlaw plots and would then increase $\alpha$. Without any evidence for this being the case, we cannot say if the remaining 88\% of the total Joule heating is in the form of small events. If they were we would most likely see a powerlaw index being close to or above two since most of the energy would then be delivered through small events. We are at the moment considering paths to establish this. 

The time evolution of the heating events is another unknown which is outside the scope of this work. What we analyse is the fingerprints of heating events at a specific time. If we do not take into account the lifetime of the events. Several authors have reported lifetimes of small and large events not being the same or even the increase and decrease of heating event being dissimilar (\citetads{1993ApJ...412..401L}; \citetads{2006SoPh..234...41K}; \citetads{2008ApJ...682..654M}; \citetads{2008ApJ...677.1385C}; \citetads{2011SSRv..159..263H}). If this is the case, that would also be able to change the powerlaws extracted from observations. 

We are able to extract information about the released energy rate and volume ranges of identified events. The smallest events we identify is significantly smaller than the lower limit for what can be observed. Even though we have much more information available through our numerical model, it is interesting how similar the problems of identifying events are from an observational and numerical stand point. The problems arise due to different limitations, but leads to the same problems. 

Our identification method is a first step towards finding the powerlaw exponent of the distribution of heating events. The exponents we report on here seems to be only lower limits, and in future work we will attempt to remove some of the problems mentioned, by looking at time evolution and the observational effects of the identified heating events. 

\bibliographystyle{aa}
\bibliography{Bibliography}{}	

\begin{thebibliography}{43}
\expandafter\ifx\csname natexlab\endcsname\relax\def\natexlab#1{#1}\fi

\bibitem[{{Aschwanden}(2015)}]{2015ApJ...814...19A}
{Aschwanden}, M.~J. 2015, \apj, 814, 19

\bibitem[{{Aschwanden} {et~al.}(2014){Aschwanden}, {Crosby}, {Dimitropoulou},
  {Georgoulis}, {Hergarten}, {McAteer}, {Milovanov}, {Mineshige}, {Morales},
  {Nishizuka}, {Pruessner}, {Sanchez}, {Sharma}, {Strugarek}, \&
  {Uritsky}}]{2014SSRv..tmp...29A}
{Aschwanden}, M.~J., {Crosby}, N.~B., {Dimitropoulou}, M., {et~al.} 2014, \ssr
  [\eprint[arXiv]{1403.6528}]

\bibitem[{{Aschwanden} \& {Shimizu}(2013)}]{2013ApJ...776..132A}
{Aschwanden}, M.~J. \& {Shimizu}, T. 2013, \apj, 776, 132

\bibitem[{{Bak} {et~al.}(1988){Bak}, {Tang}, \&
  {Wiesenfeld}}]{1988PhRvA..38..364B}
{Bak}, P., {Tang}, C., \& {Wiesenfeld}, K. 1988, \pra, 38, 364

\bibitem[{{Benz} \& {Krucker}(2002)}]{2002ApJ...568..413B}
{Benz}, A.~O. \& {Krucker}, S. 2002, \apj, 568, 413

\bibitem[{{Biskamp}(1986)}]{1986mrt..conf...19B}
{Biskamp}, D. 1986, in Magnetic Reconnection and Turbulence, ed. M.~A.
  {Dubois}, D.~{Gr{\'e}sellon}, \& M.~N. {Bussac}, 19

\bibitem[{{Carlsson} {et~al.}(2007){Carlsson}, {Hansteen}, {de Pontieu},
  {McIntosh}, {Tarbell}, {Shine}, {Tsuneta}, {Katsukawa}, {Ichimoto},
  {Suematsu}, {Shimizu}, \& {Nagata}}]{2007PASJ...59S.663C}
{Carlsson}, M., {Hansteen}, V.~H., {de Pontieu}, B., {et~al.} 2007, \pasj, 59,
  S663

\bibitem[{{Carlsson} {et~al.}(2016){Carlsson}, {Hansteen}, {Gudiksen},
  {Leenaarts}, \& {De Pontieu}}]{Carlsson2016}
{Carlsson}, M., {Hansteen}, V.~H., {Gudiksen}, B.~V., {Leenaarts}, J., \& {De
  Pontieu}, B. 2016, \aap, 585, A4

\bibitem[{{Carlsson} \& {Leenaarts}(2012)}]{CarlssonLeenaarts2012}
{Carlsson}, M. \& {Leenaarts}, J. 2012, \aap, 539, A39

\bibitem[{{Carlsson} \& {Stein}(2002)}]{2002ApJ...572..626C}
{Carlsson}, M. \& {Stein}, R.~F. 2002, \apj, 572, 626

\bibitem[{{Christe} {et~al.}(2008){Christe}, {Hannah}, {Krucker}, {McTiernan},
  \& {Lin}}]{2008ApJ...677.1385C}
{Christe}, S., {Hannah}, I.~G., {Krucker}, S., {McTiernan}, J., \& {Lin}, R.~P.
  2008, \apj, 677, 1385

\bibitem[{{Fokker}(1967)}]{1967SoPh....2..316F}
{Fokker}, A.~D. 1967, \solphys, 2, 316

\bibitem[{{Galsgaard} \& {Nordlund}(1996)}]{1996JGR...10113445G}
{Galsgaard}, K. \& {Nordlund}, {\AA}. 1996, \jgr, 101, 13445

\bibitem[{{Georgoulis} \& {Vlahos}(1996)}]{1996ApJ...469L.135G}
{Georgoulis}, M.~K. \& {Vlahos}, L. 1996, \apjl, 469, L135

\bibitem[{{Gold}(1964)}]{Goldbook}
{Gold}, T. 1964, in NASA SP-50, ed. W.~Hess, 389

\bibitem[{{Gudiksen} {et~al.}(2011){Gudiksen}, {Carlsson}, {Hansteen}, {Hayek},
  {Leenaarts}, \& {Mart{\'{\i}}nez-Sykora}}]{Gudiksen2011}
{Gudiksen}, B.~V., {Carlsson}, M., {Hansteen}, V.~H., {et~al.} 2011, \aap, 531,
  A154

\bibitem[{{Gudiksen} \& {Nordlund}(2002)}]{2002ApJ...572L.113G}
{Gudiksen}, B.~V. \& {Nordlund}, {\AA}. 2002, \apjl, 572, L113

\bibitem[{{Gudiksen} \& {Nordlund}(2005)}]{2005ApJ...618.1020G}
{Gudiksen}, B.~V. \& {Nordlund}, {\AA}. 2005, Apj, 618, 1020

\bibitem[{Gul-Mohammed {et~al.}(2014)Gul-Mohammed, Arganda-Carreras, Andrey,
  Galy, \& Boudier}]{Gul-Mohammed2014}
Gul-Mohammed, J., Arganda-Carreras, I., Andrey, P., Galy, V., \& Boudier, T.
  2014, BMC Bioinformatics, 15, 9

\bibitem[{{Hannah} {et~al.}(2011){Hannah}, {Hudson}, {Battaglia}, {Christe},
  {Ka{\v s}parov{\'a}}, {Krucker}, {Kundu}, \& {Veronig}}]{2011SSRv..159..263H}
{Hannah}, I.~G., {Hudson}, H.~S., {Battaglia}, M., {et~al.} 2011, \ssr, 159,
  263

\bibitem[{{Hansteen} {et~al.}(2015){Hansteen}, {Guerreiro}, {De Pontieu}, \&
  {Carlsson}}]{2015ApJ...811..106H}
{Hansteen}, V., {Guerreiro}, N., {De Pontieu}, B., \& {Carlsson}, M. 2015,
  \apj, 811, 106

\bibitem[{{Hoshen} \& {Kopelman}(1976)}]{1976PhRvB..14.3438H}
{Hoshen}, J. \& {Kopelman}, R. 1976, \prb, 14, 3438

\bibitem[{{Hudson}(1991)}]{1991SoPh..133..357H}
{Hudson}, H.~S. 1991, \solphys, 133, 357

\bibitem[{{Janvier} {et~al.}(2014){Janvier}, {Aulanier}, {Bommier},
  {Schmieder}, {D{\'e}moulin}, \& {Pariat}}]{2014ApJ...788...60J}
{Janvier}, M., {Aulanier}, G., {Bommier}, V., {et~al.} 2014, \apj, 788, 60

\bibitem[{Janvier {et~al.}(2013)Janvier, Aulanier, Pariat, \&
  Demoulin}]{Janvier2013}
Janvier, M., Aulanier, G., Pariat, E., \& Demoulin, P. 2013, Astronomy {\&}
  Astrophysics, 555, A77

\bibitem[{{Klimchuk}(2006)}]{2006SoPh..234...41K}
{Klimchuk}, J.~A. 2006, \solphys, 234, 41

\bibitem[{{Lee} {et~al.}(1993){Lee}, {Petrosian}, \&
  {McTiernan}}]{1993ApJ...412..401L}
{Lee}, T.~T., {Petrosian}, V., \& {McTiernan}, J.~M. 1993, \apj, 412, 401

\bibitem[{{Leenaarts} {et~al.}(2009){Leenaarts}, {Carlsson}, {Hansteen}, \&
  {Rouppe van der Voort}}]{2009ApJ...694L.128L}
{Leenaarts}, J., {Carlsson}, M., {Hansteen}, V., \& {Rouppe van der Voort}, L.
  2009, \apjl, 694, L128

\bibitem[{{Low}(1990)}]{Low1990}
{Low}, B.~C. 1990, \araa, 28, 491

\bibitem[{{Lu} \& {Hamilton}(1991)}]{1991ApJ...380L..89L}
{Lu}, E.~T. \& {Hamilton}, R.~J. 1991, \apjl, 380, L89

\bibitem[{{Morales} \& {Charbonneau}(2008)}]{2008ApJ...682..654M}
{Morales}, L. \& {Charbonneau}, P. 2008, \apj, 682, 654

\bibitem[{{Morales} \& {Charbonneau}(2009)}]{2009ApJ...698.1893M}
{Morales}, L. \& {Charbonneau}, P. 2009, \apj, 698, 1893

\bibitem[{Ollion {et~al.}(2013)Ollion, Cochennec, Loll, Escud{\'{e}}, \&
  Boudier}]{Ollion2013}
Ollion, J., Cochennec, J., Loll, F., Escud{\'{e}}, C., \& Boudier, T. 2013,
  Bioinformatics (Oxford, England), 29, 1840

\bibitem[{{Parker}(1972)}]{1972ApJ...174..499P}
{Parker}, E.~N. 1972, \apj, 174, 499

\bibitem[{{Parker}(1983{\natexlab{a}})}]{1983ApJ...264..642P}
{Parker}, E.~N. 1983{\natexlab{a}}, \apj, 264, 642

\bibitem[{{Parker}(1983{\natexlab{b}})}]{1983ApJ...264..635P}
{Parker}, E.~N. 1983{\natexlab{b}}, \apj, 264, 635

\bibitem[{{Parker}(1988)}]{1988ApJ...330..474P}
{Parker}, E.~N. 1988, \apj, 330, 474

\bibitem[{{Parnell} \& {Jupp}(2000)}]{2000ApJ...529..554P}
{Parnell}, C.~E. \& {Jupp}, P.~E. 2000, \apj, 529, 554

\bibitem[{{Priest} \& {Forbes}(2002)}]{2002A&ARv..10..313P}
{Priest}, E.~R. \& {Forbes}, T.~G. 2002, \aapr, 10, 313

\bibitem[{{Priest} \& {Pontin}(2009)}]{2009PhPl...16l2101P}
{Priest}, E.~R. \& {Pontin}, D.~I. 2009, Physics of Plasmas, 16, 122101

\bibitem[{{Priest} \& {Titov}(1996)}]{1996RSPTA.354.2951P}
{Priest}, E.~R. \& {Titov}, V.~S. 1996, Philosophical Transactions of the Royal
  Society of London Series A, 354, 2951

\bibitem[{{Scholer}(1989)}]{1989JGR....94.8805S}
{Scholer}, M. 1989, \jgr, 94, 8805

\bibitem[{{Spitzer}(1962)}]{Spitzerbook}
{Spitzer}, L. 1962, {Physics of Fully Ionized Gases}

\end{thebibliography}
\end{document}